\documentstyle[12pt]{article}
\begin{document}
\def\doublespaced{\baselineskip=\normalbaselineskip\multiply
    \baselineskip by 150\divide\baselineskip by 100}
\doublespaced
\pagenumbering{arabic}
%
\begin{titlepage}
\begin{flushright}
{ CTEQ/408 \\
MSUHEP-50319 \\
April, 1995 }
\end{flushright}
\vspace{0.4cm}
\begin{center}
\large
{\bf Effects of QCD Resummation on Distributions of \\
 Leptons from the Decay of Electroweak Vector Bosons }
\end{center}
\vspace{0.4cm}
\begin{center}
{\bf Csaba Bal\'azs~$^{(a)}$,~~~Jianwei Qiu~$^{(b)}$ ~~~
and~~~C.-P. Yuan~$^{(a)}$}
\end{center}
\vspace{0.2cm}
\begin{center}
{(a) Department of Physics and Astronomy,
Michigan State University \\
East Lansing, MI 48824, U.S.A. \\
(b) Department of Physics and Astronomy,
Iowa State University \\
Ames, IA 50011, U.S.A.
}
\end{center}
\vspace{0.4cm}
\raggedbottom
\setcounter{page}{1}
\relax

\begin{abstract}
\noindent
We study the distributions of leptons
from the decay of electroweak vector bosons produced in hadron collisions.
The effects of the initial state multiple soft-gluon emission,
using the Collins--Soper resummation formalism, are included.
The resummed results are compared with the next-to-leading-order results
for the distributions of the transverse momentum, rapidity asymmetry,
and azimuthal angle of the decay leptons.
\end{abstract}
\vspace{1.in}
\end{titlepage}
\newpage
\section{Introduction}
\indent

With the discovery of the top quark \cite{top},
the electroweak symmetry breaking mechanism
remains one of the major mysteries of particle physics
today.
Unfortunately, the precision low energy data have told us very little
about the scalar sector, i.e. the electroweak symmetry breaking sector,
of the Standard Model (SM).
{}From CERN LEP data we learned that the mass $m_H$
of the Higgs boson has to be larger than about 60\,GeV \cite{databook}.
However, the precision low energy data do not exclude the possibility
for $m_H$ to be in the order of 1\,TeV.
At the moment, one of the largest theoretical errors
in analyzing radiative corrections to low energy data
comes from the prediction of the
fine structure constant $\alpha$ evaluated at the $Z$-boson mass
scale, due to the less precise low energy
$e^-e^+ \rightarrow {\rm hadrons}$ data \cite{slacpr}.
Without further improvement in the determination of
$\alpha(M_Z)$,
a more precise way to test the SM is to have a better measurement
of $M_W$.
If $M_W$ is measured within 40\,MeV
and the mass of the top quark within 4\,GeV, then $m_H$
can be constrained within a couple of hundred GeV \cite{mhrad}.
To reach such an accuracy in the measurement of $M_W$  at
hadron colliders, we have to know the kinematics of the
$W^\pm$-boson well.
Since $W^+$ decays into a charged lepton $\ell^+$
and a neutrino $\nu_\ell$,
the kinematics of the $W^+$ cannot be accurately known
because of the missing momentum carried by $\nu_\ell$.
It is therefore desirable to have
a good prediction on the kinematics of $\ell^\pm$ from
the decay of $W^\pm$.

The well established fact that the transverse momentum $Q_T$ distribution
of the $W^\pm$-boson cannot be described by the
next-to-leading-order (NLO) perturbative
calculation in low $Q_T$ region \cite{cernw} implies that
the transverse momentum $p_T^\ell$ of the lepton $\ell^{\pm}$ cannot
be accurately predicted by the NLO calculation,
especially for $p_T^\ell \sim M_W/2$ (mostly with low $Q_T$)
where the data dominate.
We must resum the effects of the initial state
multiple soft-gluon emission
to predict the distributions of the leptons from the
decay of the vector boson $V$ ($=\gamma, \,\, W^\pm$ or $Z$)
produced in hadron collisions.\footnote{
The analytic results presented in the paper also apply to the
non-standard weak gauge boson(s) (such as $Z'$) present in any extended
gauge theory.}
In this paper, we adopt the Collins--Soper formalism \cite{collins},
and closely follow the notation used in Ref.~\cite{sterman}
to resum the multiple soft-gluon effects to the transverse momentum,
rapidity asymmetry, and azimuthal angle distributions of the decay
product leptons.

\section{The Resummation Formalism}
\indent

To obtain the resummed results, we use dimensional
regularization to regulate the infrared (IR)
divergencies, and adopt  the canonical-$\gamma_5$ prescription
to calculate the anti-symmetric part of the matrix element
in $n$-dimensional space-time.\footnote{
In this prescription, $\gamma_5$ anticommutes with other
$\gamma$'s in the first four
dimensions and commutes in others \cite{gamma5c,twoloopf3}.
}
The infrared-anomalous
contribution arising from using the canonical-$\gamma_5$ prescription
was carefully handled by
applying the procedures outlined in Ref.~\cite{mirkes} for
calculating both the virtual and the real diagrams.\footnote{
In  Ref.~\cite{mirkes} the authors calculated the anti-symmetric structure
function $F_3$ for deep-inelastic scattering.
}

The kinematics of the vector boson $V$ (real or virtual)
can be expressed in the terms of its mass $Q$, rapidity $y$,
transverse momentum $Q_T$, and azimuthal angle $\phi_V$, measured
in the laboratory frame.
The kinematics of the leptons from the decay of the vector boson
can be described by the polar angle $\theta$ and the azimuthal angle $\phi$,
defined in the Collins-Soper frame \cite{CSFrame},
which is a special rest frame of the $V$-boson \cite{lamtung}.
The four-momentum of the decay product
fermion in the lab frame is\footnote{
Our convention is that $q^\mu=(q^0,q^1,q^2,q^3)$, $Q=\sqrt{q^2}$,
and $Q_T=\sqrt{(q^1)^2+(q^2)^2}$. The total anti-symmetric tensor
$\epsilon^{0123}=-1$.
The proton beam direction is assigned to be the positive z-axis.
}
  \begin{eqnarray}
  L^\mu = {Q \over 2} \left(
  {q^\mu \over Q} +
  \sin{\theta}\cos{\phi} ~ X^\mu +
  \sin{\theta}\sin{\phi} ~ Y^\mu +
  \cos{\theta} ~ Z^\mu \right) ,
  \label{LmuV}
  \end{eqnarray}
where
  \begin{eqnarray} &&
  q^\mu =
  ( M_T \cosh{y}, \,
    Q_T \cos{\phi}, \, Q_T \sin{\phi}, \,
    M_T \sinh{y} ) ,
  \nonumber \\ &&
  X^\mu = -{Q \over Q_T M_T}
  \left( q_+ n^\mu + q_- {\bar n}^\mu - {M_T^2 \over Q^2} q^\mu \right) ,
  \nonumber \\ &&
  Z^\mu = {1 \over M_T}
  \left( q_+ n^\mu - q_- {\bar n}^\mu \right) ,
  \nonumber \\ &&
  Y^\mu = \varepsilon^{\mu \nu \alpha \beta}
  {q_\nu \over Q} Z_\alpha X_\beta .
  \label{QXYZ}
  \end{eqnarray}
Here,
$q_\pm = {1 \over \sqrt{2}}(q^0 \pm q^3)$,
$M_T=\sqrt{Q^2+Q^2_T}$,
$n^\nu={1 \over \sqrt{2}}(1,0,0,1)$ and
${\bar n}^\nu={1 \over \sqrt{2}}(1,0,0,-1)$.

To obtain the fully differential cross section of the vector boson
production and decay for all values of $Q_T$,
we need the resummation formula
\cite{sterman}:
  \begin{eqnarray} & &
  \left( { d \sigma(AB \rightarrow V(\rightarrow {l {\bar l'}}) X )
  \over dQ^2 \, dy \, dQ^2_T \, d\phi_V \, d\cos{\theta} \, d\phi}
  \right)_{res} =
  {1 \over 96 \pi^2 S} \,
  {Q^2 \over (Q^2 - M_V^2)^2 + M_V^2 \Gamma_V^2}
  \nonumber \\  & & ~~
  \times \bigg\{ {1\over (2 \pi)^2}
  \int_{}^{} d^2 b \, e^{i {\vec q_T} \cdot {\vec b}} \,
  \sum_{j,k}{\widetilde{W}_{jk} (b_*,Q,x_A,x_B, \theta, \phi)} \,
  F^{NP}_{jk} (b,Q,x_A,x_B)
  \nonumber \\  & & ~~~~
  + ~ Y(Q_T,Q,x_A,x_B, \theta, \phi) \bigg\}.
  \label{ResFor}
  \end{eqnarray}
Here $\widetilde{W}_{jk}$ is
  \begin{eqnarray} &&
  \widetilde{W}_{jk} (b,Q,x_A,x_B, \theta, \phi)  =
  \exp \left\{ -S(b,Q) \right\} \mid V_{jk} \mid^2
  \nonumber \\ && ~
  \times \left\{ \left[
  \left( C_{ja} \otimes f_{a/A} \right) (x_A) ~
  \left( C_{{\bar k}b} \otimes f_{b/B} \right) (x_B) +
  \left( C_{{\bar k}a} \otimes f_{a/A} \right) (x_A) ~
  \left( C_{jb} \otimes f_{b/B} \right) (x_B) \right] \right.
  \nonumber \\ && ~~~~~
  \times (g_L^2 + g_R^2) (f_L^2 + f_R^2) (1 + \cos^2\theta)
  \nonumber \\ && ~~~
  + \left[
  \left( C_{ja} \otimes f_{a/A} \right) (x_A) ~
  \left( C_{{\bar k}b} \otimes f_{b/B} \right) (x_B) -
  \left( C_{{\bar k}a} \otimes f_{a/A} \right) (x_A) ~
  \left( C_{jb} \otimes f_{b/B} \right) (x_B) \right]
  \nonumber \\ && ~~~~~ \left.
  \times (g_L^2 - g_R^2) (f_L^2 - f_R^2) (2 \cos\theta) \right\} ,
  \label{WTwi}
  \end{eqnarray}
where $\otimes$ denotes the convolution and is defined by
  \begin{eqnarray} & &
  \left( C_{ja} \otimes f_{a/A} \right) (x_A) =
  \int_{x_A}^{1} {d \xi_A \over \xi_A} \,
  f_{a/A}\left( \xi_A, \mu \right) ~
  C_{ja}\left( {x_A \over \xi_A}, b, \mu \right),
  \label{Convol}
  \end{eqnarray}
and the $V_{jk}$ coefficients are given by
  \begin{eqnarray}
  V_{jk} = \cases{
  {\rm Cabibbo-Kobayashi-Maskawa~matrix~elements}
 & for $V = W^\pm $ \cr
  $$\delta_{jk}$$             & for $V = Z^0,\gamma$ }.
  \end{eqnarray}
In the above expressions $j$ represents quark flavors
and $\bar{k}$ stands for anti-quark flavors.
The dummy indices $a$ and $b$
are meant to sum over quarks and anti-quarks or gluons.
Summation on these double indices is implied.

The Sudakov form factor $S(b,Q)$ is defined as
  \begin{eqnarray} & &
  S(b,Q) =
  \int_{b_0^2/b^2}^{Q^2}
  {d {\bar \mu}^2\over {\bar \mu}^2}
       \left[ \ln\left({Q^2\over {\bar \mu}^2}\right)
        A\big(\alpha_s({\bar \mu})\big) +
        B\big(\alpha_s({\bar \mu})\big)
       \right].
  \label{SudExp}
  \end{eqnarray}
The $A$, $B$ functions and the  Wilson coefficients $C_{ja}$, etc.,
were given in Ref.~\cite{sterman}.
After fixing the renormalization constants
$C_1 \equiv b_0 = 2 e^{-\gamma_E}$ and $C_2 = 1$,
one can obtain $A^{(1)}$, $B^{(1)}$, $A^{(2)}$ and $B^{(2)}$
from the Eqs.\,(3.19) to (3.22) of Ref.~\cite{sterman}.\footnote{
For instance, we obtain $A^{(1)} = 4/3$ and $B^{(1)} = -2$.
In our numerical results we also include
$A^{(2)}$ and  $B^{(2)}$.
}
($\gamma_E$ is the Euler constant.)
After choosing $\mu$ such that $\mu b \equiv C_3 = 2 e^{-\gamma_E}$,
the Wilson coefficients $C^{(i)}_{ja}$ for the parity-conserving part
of the resummed result
are greatly simplified from the Eqs. (3.23) to (3.26) of
Ref.~\cite{sterman} as
 \begin{eqnarray}
  C^{(1)}_{jk} =
  \delta_{jk} \left\{ {2 \over 3} (1 - z) +
  {1 \over 3} (\pi^2 - 8) \, \delta(1 - z)\right\}
  ~~~ {\rm and} ~~~
  C^{(1)}_{jg} = {1 \over 2} z (1 - z) .
  \label{C1jkg}
  \end{eqnarray}
Following the procedures given in Ref.~\cite{mirkes} for handling the
$\gamma_5$'s in $n$-dimensional space-time,
we find that the same Wilson coefficients $C^{(i)}_{ja}$ also apply to the
parity-violating part of the resummed result.

In Eq.~(\ref{ResFor}), the impact parameter $b$ is to be integrated
from 0 to $\infty$.
However, for $b \ge b_{max}$, which corresponds to an energy scale
less than $1/b_{max}$, the
QCD coupling $\alpha_s$ becomes so large that a perturbative
calculation is no longer reliable.\footnote{
We use $b_{max}=0.5\,{\rm GeV}^{-1}$ in our calculation.
}
Hence, the non-perturbative function
$F^{NP}$ is needed in the formalism, and generally
has the  structure
  \begin{eqnarray}
  F^{NP}_{jk} (b,Q,Q_0,x_A,x_B) =
  \exp \left[-\ln \left( Q^2\over Q^2_0 \right) h_1(b)
    -h_{j/A}(x_A,b)-h_{{\bar k}/B}(x_B,b)\right] ,
  \label{FNPh}
  \end{eqnarray}
where $h_1$, $h_{j/A}$ and $h_{{\bar k}/B}$ cannot be calculated using
perturbation theory, so they must be measured experimentally.
Furthermore, $\widetilde{W}$ is evaluated at $b_*$, with
  \begin{eqnarray}
  b_* = {b \over \sqrt{1+(b/b_{max})^2} }
  \label{bStar}
  \end{eqnarray}
such that $b_*$ never exceeds $b_{max}$.

The $Y$-term in Eq.~(\ref{ResFor}) is defined as
  \begin{eqnarray}
  Y(Q_T,Q,x_A,x_B,\theta,\phi) =
  \int_{x_A}^{1} {d \xi_A \over \xi_A}
  \int_{x_B}^{1} {d \xi_B \over \xi_B}
  \sum_{N=1}^{\infty} \left[{\alpha_s(Q) \over \pi} \right]^N
  \nonumber \\
  \times f_{a/A}(\xi_A;Q) \, R_{ab}^{(N)} (Q_T,Q,z_A,z_B,\theta,\phi,Q)
  \, f_{b/B}(\xi_B;Q) ,
  \label{RegPc}
  \end{eqnarray}
in which the functions $R_{ab}^{(N)}$
only contain contributions which are less singular than
{\hbox{$Q_T^{-2} \times $ (logs~or~1)}} as $Q_T \rightarrow 0$.
Their explicit expressions for
${\bar{\rm p} {\rm p}}\rightarrow V( \rightarrow l {\bar l'}) X$
are given in the Appendix.

\section{Numerical Results}
\indent

In this paper, we only give our numerical results for
${\bar{\rm p} {\rm p}} \rightarrow W^+ ( \rightarrow \ell^+ \nu_\ell ) X $
at the Fermilab Tevatron with $\sqrt{S} = 1.8\,$TeV.
The CTEQ3M parton distribution functions (PDF's)
are used along with the non-perturbative function~\cite{glenn}
  \begin{eqnarray}
  F^{NP} (b,Q,Q_0,x_A,x_B) = {\rm exp}
  \left\{- g_1 b^2 - g_2 b^2 \ln\left( {Q \over 2 Q_0} \right) -
  g_1 g_3 b \ln{(100 x_A x_B)} \right\},
  \label{FNPg}
  \end{eqnarray}
where $g_1 = 0.11\,{\rm GeV}^2$,
$g_2 = 0.58\,{\rm GeV}^2$, $g_3 = -1.5\,{\rm GeV}^{-1}$
and $Q_0 = 1.6\,{\rm GeV}$.\footnote{
These values were fit for CTEQ2M PDF, and in principle should be refit
for CTEQ3M PDF.
}
To consistently compare the distributions of the leptons in  NLO and
resummed calculations, we have used exactly the same PDF's,
QCD and electroweak parameters, etc.,
for calculating the NLO results.\footnote{
Our NLO results agree with those in Ref.~\cite{giele}.}
Furthermore, we have applied the kinematic cuts
$p_T^{\ell} > 25\,$GeV, ${\not{\hbox{\kern-4pt $E_T$}}} > 25\,$GeV,
and $Q_T < 20\,$GeV.
These cuts are similar to those applied by the CDF group
in the measurement of the asymmetry in the lepton rapidity
distribution from $W$-boson decays.\footnote{
The requirement of $Q_T < 20\,$GeV in our calculation
is approximately equivalent to cutting out the events
in which the transverse momentum of the net hadronic activities
reconstructed from the calorimeter cells within the
pseudo-rapidity range of $\pm 3.5$ is larger than $20\,$GeV.
}

\begin{figure}
\let\picnaturalsize=N
\def\picsize{5.0in}
\def\picfilename{FigpTe.eps}
\ifx\nopictures Y\else{\ifx\epsfloaded Y\else\input epsf \fi
\let\epsfloaded=Y
\centerline{\ifx\picnaturalsize N\epsfxsize
\picsize\fi\epsfbox{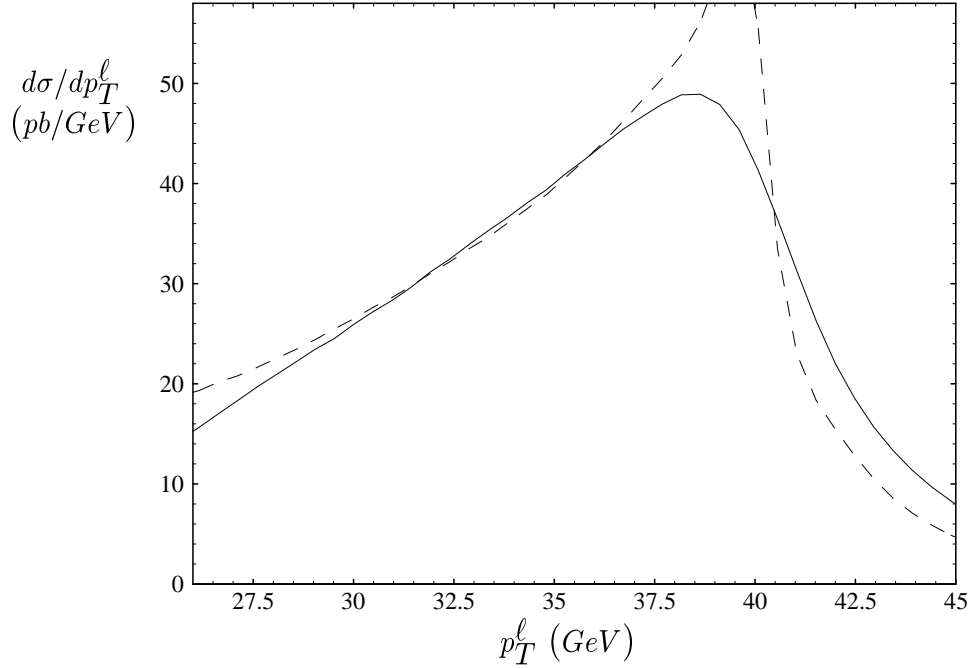}}}\fi
\caption{Transverse momentum distribution of the charged lepton $p_T^\ell$
for NLO (dashed) and resummed (solid) calculations.
Resumming the initial state multiple soft-gluon emission has
the typical effect of smoothening and widening the Jacobian peak
(at $p_T^\ell = M_W/2$).
The NLO distribution is singular and ill-defined near $M_W/2$.
}
\label{fig1}
\end{figure}

The transverse momentum distributions of the charged
lepton $p_T^\ell$ are shown in Fig.~\ref{fig1} for NLO and resummed
calculations. We note that in these results a Breit-Wigner resonant width
has been included, cf. Eq.~(\ref{ResFor}).
In the vicinity of $p_T^\ell = M_W/2$ (``Jacobian peak'')
the NLO calculation is ill-defined
because its amplitude blows up as $Q_T \rightarrow 0$.
The effect of the
initial state multiple soft-gluon emission on the distribution of
$p_T^\ell$ is to widen and smoothen the ``Jacobian peak'' and
therefore make it more challenging to accurately extract
$M_W$ from the $p_T^\ell$ distribution.

Recently, the Fermilab CDF group measured the asymmetry
${\cal A}_{y^\ell}$ in the rapidity $y^\ell$
distribution of the charged lepton ($\ell$) from the decay of
$W^+ \rightarrow  \ell^+ \nu_{\ell}$
(or $W^- \rightarrow  \ell^- \bar{\nu}_{\ell}$) \cite{CDF} and proved
this measurement to be particularly sensitive to the slope of
the ratio of $u$- to $d$-quark  parton densities
inside the proton \cite{berger,cteq3}.
\begin{figure}
\let\picnaturalsize=N
\def\picsize{4.5in}
\def\picfilename{FigLCA.eps}
\ifx\nopictures Y\else{\ifx\epsfloaded Y\else\input epsf \fi
\let\epsfloaded=Y
\centerline{\ifx\picnaturalsize N\epsfxsize
\picsize\fi\epsfbox{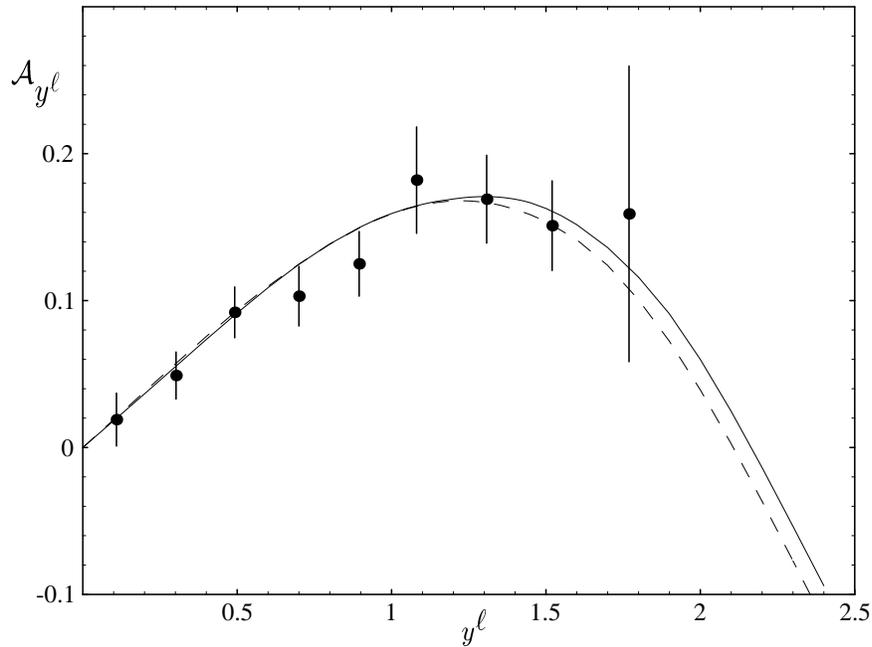}}}\fi
\caption{
The asymmetry ${\cal A}_{y^\ell}$ in the lepton rapidity distribution
 as a function of ${y^\ell}$ for
NLO (dashed) and resummed (solid) calculations.
They differ the most
in the large rapidity region ($|y_{\ell^+}| > 1$).
The experimental data were obtained from Ref.~[15]
}
\label{fig2}
\end{figure}
\begin{figure}
\let\picnaturalsize=N
\def\picsize{4.5in}
\def\picfilename{FigDelPhi.eps}
\ifx\nopictures Y\else{\ifx\epsfloaded Y\else\input epsf \fi
\let\epsfloaded=Y
\centerline{\ifx\picnaturalsize N\epsfxsize
\picsize\fi\epsfbox{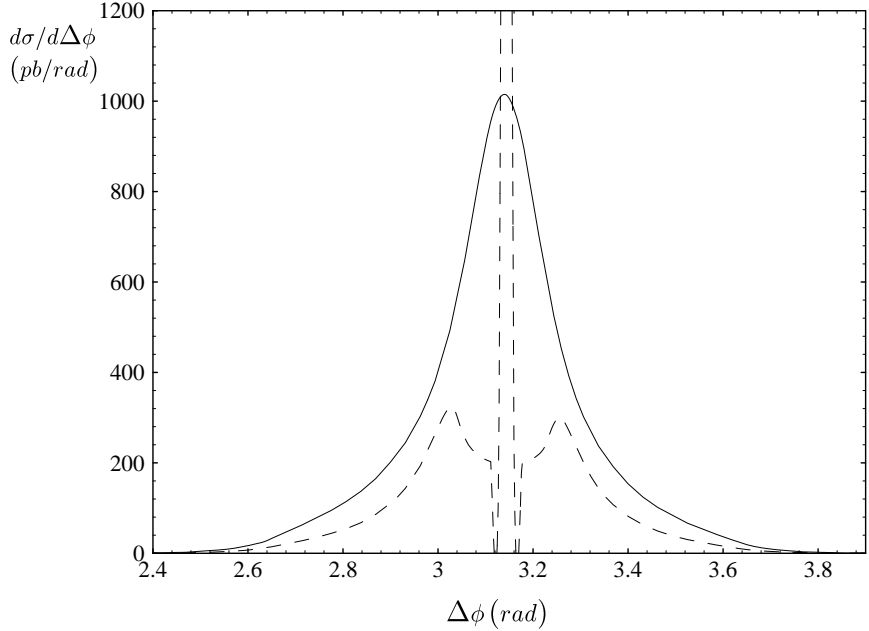}}}\fi
\caption{
The distribution of the difference in the lepton azimuthal angles
near the region $\Delta \phi \sim \pi$.
The NLO (dashed) distribution is ill-defined at $\Delta \phi = \pi$
and is arbitrary around it.
The resummed (solid) distribution
gives the correct angular correlation of the lepton pair.
The distribution has a similar peak for $\Delta \phi \sim -\pi$.
}
\label{fig3}
\end{figure}
Define the asymmetry in the lepton rapidity distribution as
  \begin{eqnarray}
  {\cal A}_{y^{\ell}} =
  {{d \sigma / d y^{\ell}} (y^{\ell} > 0) -
   {d \sigma / d y^{\ell}} (y^{\ell} < 0)
  \over
   {d \sigma / d y^{\ell}} (y^{\ell} > 0) +
   {d \sigma / d y^{\ell}} (y^{\ell} < 0) },
  \label{eq22}
  \end{eqnarray}
which is commonly known as the lepton charge asymmetry.
In Fig.~\ref{fig2}, we show ${\cal A}_{y^\ell}$ as
a function of ${y^\ell}$ for
NLO  and resummed calculations.
As indicated in the figure, they differ the most
in the large rapidity region ($|y_{\ell^+}| > 1$).
Recall that
as $Q_T \rightarrow 0$ the NLO $Q_T$ distribution becomes singular,
but the resummed result remains finite.
The rapidity $y$ distribution of the $W$-boson in the NLO calculation is
not singular, and we expect that after integrating out the
complete phase space for $Q_T$
(that is, without imposing any kinematic cuts)
the NLO and the resummed calculations should predict the same
$y$ distributions.
We have explicitly checked that this indeed is the case.
However, in Fig.~\ref{fig2}
some kinematic cuts (as described at the beginning of this section)
have been applied to our calculations.
In the rest (Collins-Soper) frame of the $W$-boson,
the decay kinematics of the lepton $\ell$ is identical for both
the NLO and the resummed calculations because the decay leptons do not
involve strong interactions. Since the $W$-bosons have
different kinematic distributions (e.g., $Q_T$ distributions)
in these two calculations, the resulting lepton kinematic
distributions (e.g., $y^\ell$ distributions)
in the laboratory frame are different.
The two $y^\ell$ distributions
differ the most in the large $y^\ell$ region,
where the typical $y$ is large,
since the effects of soft-gluon emission become more important,
close to the boundary of the phase space.

Another interesting observable to test the QCD theory
beyond the fixed-order perturbative calculation is the measurement of the
difference  in the azimuthal angles
of $l$ and ${\bar l'}$ from
the decay of $V$. In practice, this is better measured for
$Z \rightarrow \ell^+ \ell^-$. For the sake of argument, we
show in Fig.~\ref{fig3} the difference ($\Delta \phi$) in the azimuthal angles
of $\ell^+$ and $\nu_\ell$ measured in the laboratory frame
for $W^+ \rightarrow \ell^+ \nu_\ell$ and calculated in NLO and
resummed approaches. As clearly indicated, the NLO result
is ill-defined in the vicinity of $\Delta \phi \sim \pm \pi$,
where the multiple soft-gluon radiations have to be resummed
to obtain physical predictions.
Therefore, the transverse mass distributions of $\ell^+$-$\nu_\ell$
pair for NLO and resummed calculations are also different,\footnote{
The transverse mass of $\ell^+$-$\nu_\ell$ pair is
defined as
$m^{\ell \nu}_T=\sqrt{2 p^\ell_T p^\nu_T (1 - \cos \Delta \phi)}$.
}
and in principle only the resummed results can sensibly predict
the distributions of the  leptons for a precise
measurement of $M_W$.

In conclusion, we found that the distributions ($p_T^\ell$,
$y^\ell$ and $\Delta \phi$) of leptons  are different in
NLO and resummed calculations.
For a better measurement of $M_W$ and $A_{y^\ell}$,
the effects of the initial state multiple soft-gluon emission
have to be considered in hadron collisions.
The more detailed phenomenological studies
will be presented elsewhere.

\section*{ Acknowledgments }
\indent

We thank
E.L. Berger, R. Brock, G.A. Ladinsky, Wu-Ki Tung, and
the CTEQ collaboration for many invaluable discussions.
This work was supported in part by NSF under grant PHY-9309902
and by DOE under grant DE-FG02-92ER40730 and DE-FG02-87ER40731.

\vspace{0.4cm}
\section*{Appendix}
\indent

Let us define the $q \bar{q'} V$ and the
$l \bar{l'} V$ vertices, respectively, as
  \begin{eqnarray}
  i \gamma_{\mu} \left[ g_L (1 - \gamma_5) + g_R (1 + \gamma_5) \right]
  ~ ~ ~ {\rm and} ~ ~ ~
  i \gamma_{\mu} \left[ f_L (1 - \gamma_5) + f_R (1 + \gamma_5) \right] .
  \end{eqnarray}
For example, for $V = W^+, q = u$, ${\bar q'} = {\bar d}$,
$l = \nu_{e}$, and ${\bar l'} = e^+$, the couplings
$ g_L^2 = f_L^2 = G_F M_W^2/\sqrt{2}$ and $g_R^2 = f_R^2 = 0$.
($G_F$ is the Fermi constant.)
In Eq.~(\ref{RegPc}), for $N = 1$,
\begin{eqnarray} & &
  R_{ab}^{(1)} = {16 \mid V_{jk} \mid^2 \over \pi Q^2}
  \left[
  (g_L^2 + g_R^2) (f_L^2 + f_R^2) R_1^{ab} +
  (g_L^2 - g_R^2) (f_L^2 - f_R^2) R_2^{ab}
  \right] ,
\end{eqnarray}
where the coefficient functions $R^{ab}_i$ are given as follows:
\begin{eqnarray}  & &
  R_1^{j{\bar k}} =
  r^{j {\bar k}} {\cal L}_0 +
  {{\cal R_+}(t,u) \over s} \,
  \delta (s + t + u - Q^2)
  \left [ {\cal A}_0 + {\cal A}_2 + {Q \over Q_T} {\cal A}_1 \right ]
  {Q^2 \over M_T^2}
 , \nonumber \\  & &
  R_2^{j{\bar k}} =
  r^{j{\bar k}} {\cal A}_3 +
  {{\cal R_+}(t,u) \over s} \,
  \delta (s + t + u - Q^2)
  \nonumber \\  & & ~~~~~~~~~~~~ \times
  \left\{ {Q^2 \over Q_T^2} \left( {Q \over M_T} - 1 \right) {\cal A}_3 -
  {2 Q^2 \over Q_T M_T} {{\cal R_-}(t,u) \over {\cal R_+}(t,u)} {\cal A}_4
  \right\} ,
  \end{eqnarray}
\begin{eqnarray}  & &
  R_1^{gj} =
  r^{gj} {\cal L}_0 -
  {Q^2 Q_T^2 \over u M_T^2} \,
  {{\cal R_+}(u,s) \over s} \,
  \delta (s + t + u - Q^2)
  \nonumber \\  & & ~~~~~~~~~~~~ \times
  \left\{ {{\cal R_+}(u,-s) \over {\cal R_+}(u,s)}
  \left[ {\cal A}_0 + {\cal A}_2 \right] +
  {Q \over Q_T} {(Q^2 - u)^2 + {\cal R_-}(u,t) \over {\cal R_+}(u,s)}
  {\cal A}_1 \right\}
  , \nonumber \\  & &
  R_2^{gj} =
  r^{gj} {\cal A}_3 -
  {Q_T^2 \over u} ~
  {{\cal R_+}(u,s) \over s} \,
  \delta (s + t + u - Q^2)
  \nonumber \\  & & ~~~~~~~~~~~~ \times
  \left\{ {Q^2 \over Q_T^2} \left[ {Q \over M_T}
  \left( {2 u (Q^2 - s) \over {\cal R_+}(u,s)} - 1 \right) - 1 \right]
    {\cal A}_3
  \right.
  \nonumber \\  & & ~~~~~~~~~~~~~~~~~~~~
  \left.
  -{2 Q^2 \over Q_T M_T} \left[ {2 s (Q^2-s) \over {\cal R_+}(u,s)}+1\right]
    {\cal A}_4
  \right\} ,
  \nonumber
  \end{eqnarray}
with
  \begin{eqnarray} & &
  r^{j {\bar k}} = {Q^2 \over Q_T^2}
  \Bigg\{
    {{\cal R}_+ (t,u) \over s} ~ \delta (s + t + u - Q^2)  -
    2 ~ \delta (1 - z_A) ~ \delta (1 - z_B)
      \left[ \ln{\left( {Q^2 \over Q_T^2} \right)} - {3 \over 2}\right]
  \nonumber \\  & &  ~~~~~~~~~~~~~~~
    - \delta (1 - z_A) \left( {1 + z_B^2 \over 1 - z_B} \right)_+ -
      \delta (1 - z_B) \left( {1 + z_A^2 \over 1 - z_A} \right)_+
  \Bigg\} ,
  \end{eqnarray}
and
  \begin{eqnarray}
  r^{gj} = {Q^2 \over Q_T^2}
  \left\{ - {Q_T^2 \over u} ~ {{\cal R_+}(u,s) \over s} \,
  \delta (s + t + u - Q^2) -
  \left[ z_A^2 + (1 - z_A)^2 \right] \, \delta (1 - z_B)
  \right\} ,
  \end{eqnarray}
where ${\cal R_{\pm}}(t,u) = (Q^2 - t)^2 \pm (Q^2 - u)^2$.
The angular dependence is described by the functions
  \begin{eqnarray} & &
  {\cal L}_0 = 1 + \cos^2{\theta}, ~
  {\cal A}_0 = {1 \over 2} (1 - 3 \cos^2{\theta}), ~
  {\cal A}_1 = \sin{2 \theta} \cos{\phi}, ~
  {\cal A}_2 = {1 \over 2} \sin^2{\theta} \cos{2 \phi}, ~
  \nonumber \\  & &
  {\cal A}_3 = 2 \cos{\theta}, ~
  {\cal A}_4 = \sin{\theta} \cos{\phi},
  \end{eqnarray}
of which ${\cal A}_3$ and ${\cal A}_4$ are odd under parity operation.

\end{document}